\documentclass[aps,pra,twocolumn,showpacs]{revtex4}
\usepackage{times}
\usepackage{amsmath}
\usepackage{amssymb}
\usepackage{graphicx}
\usepackage{epstopdf}

\newcommand{\abs}[1]{\left|{#1}\right|}
\newcommand{\re}{\mathrm{Re}}
\newcommand{\im}{\mathrm{Im}}

\newcommand{\atan}{\mathrm{atan}}

\newcommand{\asym}{\mathcal{A}}
\newcommand{\RR}{\mathbb{R}}

\newcommand{\DD}{\mathbf{D}}
\newcommand{\PP}{\mathcal{P}}
\newcommand{\vv}{\mathbf{v}}

\newcommand{\tosc}{\tau_{\mathrm{o}}}
\newcommand{\tdamp}{\tau_{\mathrm{d}}}

\begin{document}

\title{Collective excitations and instability of an optical lattice
due to unbalanced pumping}

\author{J.\ K.\ Asb\'oth$^{1,2}$}

\author{H.\ Ritsch$^{1}$}

\author{P.\ Domokos$^{2}$}

\affiliation{$^1$Institut f\"ur theoretische Physik, Universit\"at
Innsbruck, Technikerstr.~25, A-6020 Innsbruck, Austria\\
$^2$Research Institute of Solid State Physics and Optics, Hungarian
Academy of Sciences, H-1525 Budapest P.O. Box 49, Hungary}

\begin{abstract} 
  We solve self-consistently the coupled equations of motion for
  trapped particles and the field of a one-dimensional optical
  lattice. Optomechanical coupling creates long-range interaction
  between the particles, whose nature depends crucially on the
  relative power of the pump beams. For asymmetric pumping, traveling
  density wave-like collective oscillations arise in the lattice, even
  in the overdamped limit. Increasing the lattice size or pump
  asymmetry these waves can destabilize the lattice.
\end{abstract}

\pacs{32.80.Lg,42.65.Sf,63.22.+m,71.36.+c}

\maketitle

Optical lattices (OL) are perfectly periodic arrays of particles
trapped by the standing wave interference pattern of several laser
beams. They have important applications as model systems for solid
state physics as well as for quantum information science.
The back-action of the trapped particles on the trap light is
carefully avoided in most OL experiments. However, it is known to give
rise to intriguing phenomena in related systems, e.g.,
cavity cooling \cite{cavcool}, mirror cooling \cite{mirror06}, and 
optical binding \cite{optical_binding}. For OL's this
back-action has been predicted \cite{deutsch95} and observed
\cite{birkl95,weidemuller98} to reduce the lattice constant
compared to the naive expectation.

In this Letter we consider the dynamical effects of optical
back-action in a one-dimensional OL, brought about by tuning a
hitherto neglected parameter, the asymmetry in the intensities of the
lattice beams.  Due to the back-action the trap light mediates an
\emph{interaction} between the particles, which is substantially
altered by this asymmetry. Net energy and momentum flow is induced
through the OL, relating it to crystals driven far from equilibrium,
e.g., arrays of vortices in a type-II superconductor
\cite{PhysRevLett.83.3285}, and trains of water drops dragged by oil
\cite{beatus06}.  The phonon-like traveling density waves
characteristic of these systems become the elementary excitations of
the OL as well, and can destabilize it, even in the presence of
arbitrarily strong viscous damping. The excitations arise resonantly
at specific values of the asymmetry, which allows for tuning the
dispersion relation of the lattice.  Moreover, the light-mediated
interaction in the OL is of infinite range, and thus all these effects
depend heavily on the size of the lattice. As absorption inevitably
leads to pumping asymmetry, this dynamic instability limits the size
of any near-resonant OL.

\begin{figure}
\begin{center}
  \includegraphics[width=8cm]{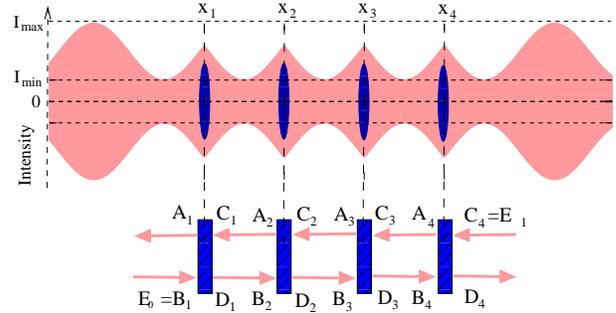}
\caption{(color online) A dipole trap created by two lasers
  of equal frequency but unequal power. The intensity (in light red),
  mirrored for better visibility, ranges between $I_{\mathrm{min}} =
  \frac{1}{2}\epsilon_0 \,c\, (\abs{E_0}-\abs{E_1})^2$ and
  $I_{\mathrm{max}} = \frac{1}{2}\epsilon_0\, c\,
  (\abs{E_0}+\abs{E_1})^2$. Trapped particles form disk-shaped clouds
  (in dark blue), and are modeled as beam splitters. Due to the pump
  asymmetry, the electric field has no nodes. Back-action of trapped particles 
  distorts the field and reduces the lattice constant.
\label{fig:system} 
}
\end{center}
\end{figure}

We consider a dipole trap formed by two counter-propagating phase
locked laser beams with frequency $\omega=c k$. The waist of the trap
is much larger than the wavelength $\lambda=2\pi/k$, so the light
field is essentially 1 dimensional along $x$.  The two beams have
unequal intensities: the electric field incident from the left is
$E(x)= E_0 e^{ikx-i\omega t}$, from the right, $E(x)= E_1
e^{-ikx-i\omega t}$, with $E_1=e^{i\phi} \sqrt{\PP} E_0$, and $\phi$ the
relative phase.  Besides the pump power ratio $\PP>1$, we
introduce another measure of the asymmetry:
$\asym=\abs{E_1/E_0}-\abs{E_0/E_1}$.  We consider particles of linear
polarizability $\alpha$ and mass $m_A$ pre-cooled down to very low
temperatures (possibly pre-trapped) and trapped by the dipole force in
the light field.  These can be atoms, the lasers being detuned to the
red of a specific transition so far that it is not saturated and
spontaneous emission can be ignored.  Alternatively, they can be
plastic beads trapped in water, as in, e.g., \cite{zemanek,Monika},
albeit of size well below $\lambda$ so that complications of Mie
scattering are avoided.  If the particles are cold enough, they gather
at the antinodes, forming $N$ disk-shaped clouds of axial size much
smaller than $\lambda$. For simplicity we assume that each cloud has
the same number of particles, and thus identical surface density
$\eta$, surface mass density $m=\eta m_A$, and dimensionless
polarizability $\zeta =k \eta \alpha/ (2 \epsilon_0)$.  The setup is
sketched in Fig.~\ref{fig:system}.

We now take the back-action of the particles on the light field into
account.  Following \cite{deutsch95}, this is achieved by solving the
scalar Helmholtz equation, with the $N$ clouds represented by
Dirac-$\delta$ distributions of linearly polarizable material,
\begin{equation}
\label{eq:Helmholtz}
(\partial_x^2  + k^2)E(x) = -2k E(x) \sum_{j=1}^{N} \zeta \delta(x-x_j).
\end{equation}
Throughout this Letter we assume $\zeta\in \mathbb{R}$, 
neglecting spontaneous emission and scattering into other transverse
modes, justified as long as the laser beams are far
detuned from any resonance of the trapped particles.  Note that
although these approximations can be relaxed by setting $\zeta\in
\mathbb{C}$, very close to resonance the reabsorption of spontaneously
emitted photons plays an important role in the dynamics
\cite{labeyrie}, and this is not easily incorporated into this model.

The solution of Eq.\ (\ref{eq:Helmholtz}) between two clouds is a
superposition of plane waves, 
$E(x_{j-1}< x < x_j)= A_{j}e^{ik(x-x_j)}+B_{j}e^{-ik(x-x_j)}
=C_{j-1}e^{ik(x-x_{j-1})}+D_{j-1}e^{-ik(x-x_{j-1})}.
$
The clouds constitute boundary conditions for the field:
\begin{subequations}
\label{eq:fit}
\begin{align}
\label{eq:fit_E}
 E(x=x_j-0) &= E(x=x_j+0);\\
\label{eq:fit_grad_E}
\partial_x E(x=x_j-0) &= \partial_x E(x=x_j+0) + 2k\zeta E(x_j).
\end{align}
\end{subequations}
This amounts to representing each
cloud as a beam splitter (BS) at $x=x_j$ with reflection and
transmission coefficients $r = i \zeta/(1-i\zeta)$ and $t =
1/(1-i\zeta)$, so that $\zeta=-ir/t$ \cite{deutsch95}.

Since $E(x)$ is not differentiable at the cloud position, as given by
Eq.~\eqref{eq:fit_grad_E} and shown on Fig.~\ref{fig:system}, we need
to calculate the dipole force on the cloud carefully. Integrating the
force over a finite cloud and then taking the Dirac-$\delta$ limit, we
obtain
\begin{equation}
\label{eq:force}
F_j = \frac{\eta \alpha}{8} \left(\partial_x E^2(x_j-0)
+ \partial_x E^2(x_j+0)\right) 
\end{equation}
for the force on a unit surface of the cloud, averaged over an optical
period. This formula can also be derived based on the amount of
momentum transferred to the cloud by the field, via the Maxwell stress
tensor, as in \cite{antonoyiannakis}.

For a {\it single} cloud at steady state, both 
$F_j=0$ and Eqs.~\eqref{eq:fit} must hold, 
which is only possible if
\begin{equation}
\label{eq:critical_zeta}
\zeta \asym < 2. 
\end{equation}
This simple equilibrium criterion can be intuitively understood in the
following way.  If $\abs{E_0}^2<\abs{E_1}^2$, more photons are
incident on the right of the BS than the left, giving a force on it.
If enough light is transmitted ($\abs{t}>\frac{1}{2}\abs{r}\asym$),
and the interference is favourable (depending on the position of
the BS), the imbalance in the outgoing number of photons is enough to
counteract this force, leading to a steady state.

For several clouds trapped by the same light, at steady state $F_j$
has to vanish on all components of the system, which with
Eqs.~(\ref{eq:fit}) and (\ref{eq:force}) implies that $E(x)$ and
$\abs{\partial_x E(x)}$ are the same to the left and right of any
component. As a result, $\abs{E^2(x)}=\abs{E_0}^2+\abs{E_1}^2 + 2 \abs{E_0
E_1}\cos(2kx-\Phi(x)) $ everywhere in the sample, the clouds only
contribute to the phase: $\Phi(x_j<x<x_{j+1}) =\sum_{l=1}^{j} \chi_l
$, the phase slip at the $l$'th cloud depending on the polarizability
$\zeta_l$ of the cloud as
\begin{equation}
\label{eq:chi}
\cos \chi_l (\zeta_l,\asym) = \frac{\sqrt{4-\zeta_l^2 \asym^2} - 
\zeta_l^2 \sqrt{\asym^2+4}}{2 (1+\zeta_l^2)}.
\end{equation}
Thus, at steady state, $\abs{A_1}=,\ldots,=\abs{A_N} =
\abs{C_1}=,\ldots,=\abs{C_N}$, and $\abs{B_1}=,\ldots,=\abs{B_N} =
\abs{D_1}=,\ldots,=\abs{D_N}$, i.e., the pump lasers fill the structure
unattanuated.

\begin{figure}
\begin{center}
  \includegraphics[angle=270,width=8cm]{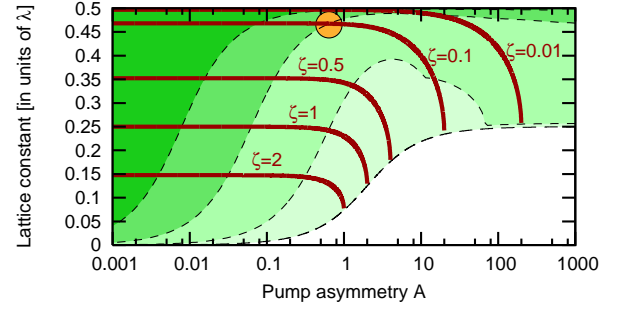}
\caption{(color online) The lattice constant as a function of the asymmetry
  is shown in thick (red) curves for $\zeta=0.01$, $\zeta=0.1$,
  $\zeta=0.5$, $\zeta=1$, $\zeta=2$. Shaded (green) areas indicate
  regions of stability (see page 4),  
  for $N\le 800$ (darkest shade), $N\le 100$,
  $N\le 10$ and $N\le 2$ (lightest shade).  The white area is
  unstable, see Eq.~\eqref{eq:critical_zeta}. 
  The orange circle marks the parameter regime of 
  Fig.~\ref{fig:waves}.
\label{fig:d_power} 
}
\end{center}
\end{figure}

Now consider the steady state of $N>1$ identical, purely dispersive
trapped clouds, with $\zeta_1=,\ldots,=\zeta_N=\zeta<2/\asym$. Since
at every cloud $\abs{C_j/B_j}-\abs{B_j/C_j}=\asym$, the phase slips
are all equal: $\chi_1=,\ldots,=\chi_N=\chi$.  Thus the equilibrium
configuration is an equidistant lattice,
$x_j=x_j^{(0)}=x_1^{(0)}+(j-1)d$. The lattice constant $d$ is clearly
independent of $N$, and a decreasing function of the phase shift
$\chi$ -- see Fig.~\ref{fig:system}, and the introduction of
\cite{deutsch95} --, explicitly
\begin{equation}
\label{eq:lattice_constant}
d = \frac\lambda{2\pi}\big(\pi-\chi(\zeta,\asym)\big).
\end{equation}
For $\asym=0$ this gives $d_{\mathrm{symm}}= \frac{\lambda}{2}
(1-2\,\atan(\zeta) /\pi)$ as in \cite{deutsch95}.  For a given $\zeta$,
increasing $\asym$ causes the phase shift $\chi$ to increase, and $d$
to be reduced, as illustrated in Fig.\ref{fig:d_power} (thick red
lines).  For $\asym>2/\zeta$, the inequality \eqref{eq:critical_zeta}
is violated, the stronger beam pushes all the particles away.  At
$\asym=2/\zeta$ the lattice constant $d$ is, remarkably, exactly half
of $d_{\mathrm{symm}}$:
\begin{equation}
d_{\mathrm{min}}(\zeta) = 
\frac\lambda{4\pi} \big(\pi - 2\,\atan \zeta \big).
\end{equation} 

The fact that an equilibrium lattice configuration exists is only
physically relevant if this equilibrium is dynamically {\it stable}. 
The dynamics of an OL is given by
\begin{equation}
\label{eq:atom_dynamics}
m \ddot{x}_j = - \mu \dot{x_j} + F_j(x_1 .. x_N),
\end{equation}
where in addition to the light-induced dipole force $F_j$ from
Eq.\eqref{eq:force}, we include viscous friction with coefficient
$\mu$ (related to the single-particle friction coefficient $\mu_A$ by
$\mu = \eta \mu_A$). For plastic beads immersed in water, $\mu$
follows from the Stokes law; for atoms in vacuum, it can represent some
laser cooling mechanism. This equation is nonlinear, as its solution
involves integrating (\ref{eq:Helmholtz}) to obtain the electric field
for the force.  We proceed by linearizing Eq.~\eqref{eq:atom_dynamics}
around an equilibrium configuration. For $\xi_j = x_j - x_j^{(0)} \ll
\lambda$ we have
\begin{equation}
\label{eq:atom_linear}
m \ddot{\xi}_j = - \mu \dot{\xi}_j + \sum_{l=1}^{N} D_{jl} \xi_l, 
\end{equation}
where the matrix $\DD$ is defined by
\begin{align}
\label{eq:Dmatrix}
\quad D_{jl} &= \frac\partial{\partial x_l} F_j(\xi_n=0,n=1..N) .
\end{align}

Stability analysis requires finding the eigenvectors of $\DD$ and
determining their dynamics. Details of this calculation are involved
and will be published elsewhere, we outline the procedure below.  The
key tool is the transfer matrix (TM) method, as used in
\cite{deutsch95}. The TM of the whole optical lattice is a product of
the TM's of a single block of the lattice, which consist of the BS
transformation followed by free propagation over length $d$. 
Since losses are neglected,
the two eigenvalues of the TM of a single block are $e^{\pm
i\Theta}$ with $\Theta\in \mathbb{C}$.  The parameter $\Theta$, 
related to the quasimomentum, is given by the solution of $\cos
\Theta = \cos kd - \zeta \sin kd$ \cite{deutsch95}. We solve this
equation explicitly and obtain the surprisingly simple result
\begin{equation}
\label{eq:theta}
\sin \Theta = 
\zeta \asym /2; \quad \pi/2<\Theta<\pi.
\end{equation}

We next apply the TM method to a perturbed OL where the $l$'th cloud
is displaced by an infinitesimal amount. The calculations lead to
explicit formulas for the matrix $\DD$ which we omit here for the sake
of brevity. Two important properties of $\DD$ must be mentioned.
First, $D_{jl}$ depends only on $l-j$: $\DD$ is Toeplitz matrix. In
particular, for $D_{jj}<0$ all clouds are trapped in identical wells.
Second, $\DD$ is not symmetric. This shows that $F_j$ is not a
conservative force: if it were, $F_j=-\partial/\partial x_j \,
V(x_1...x_N)$ would imply that $\DD$ is a Hessian matrix, symmetric by
Young's theorem. Note that reflection symmetry of the system is broken
by the pump asymmetry.

\begin{figure}
\begin{center}
\includegraphics[angle=270,width=8cm]{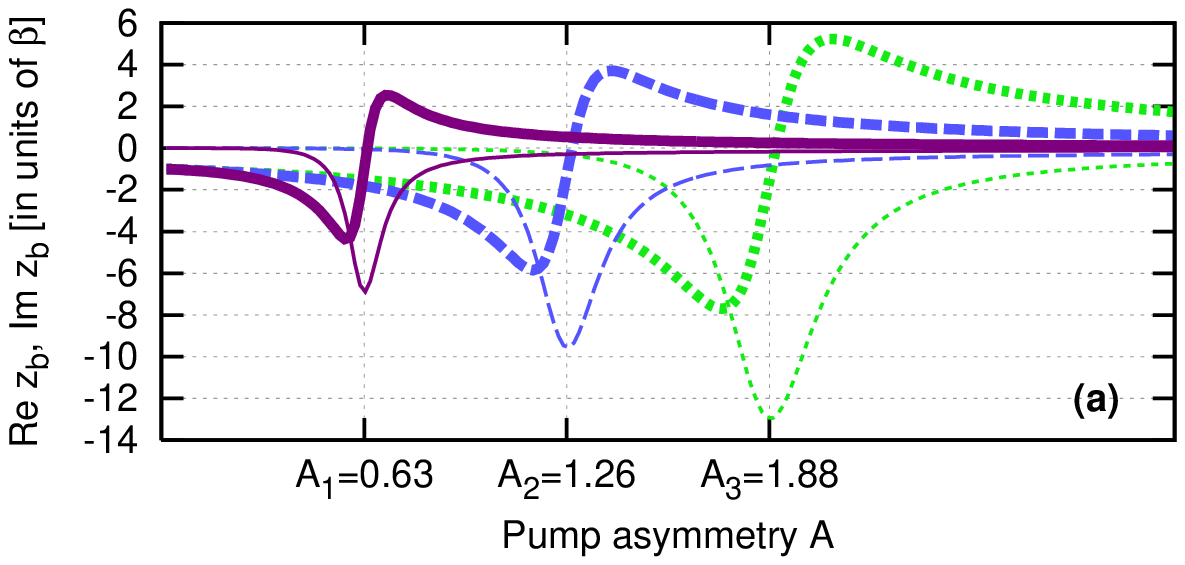}
\includegraphics[angle=270,width=8cm]{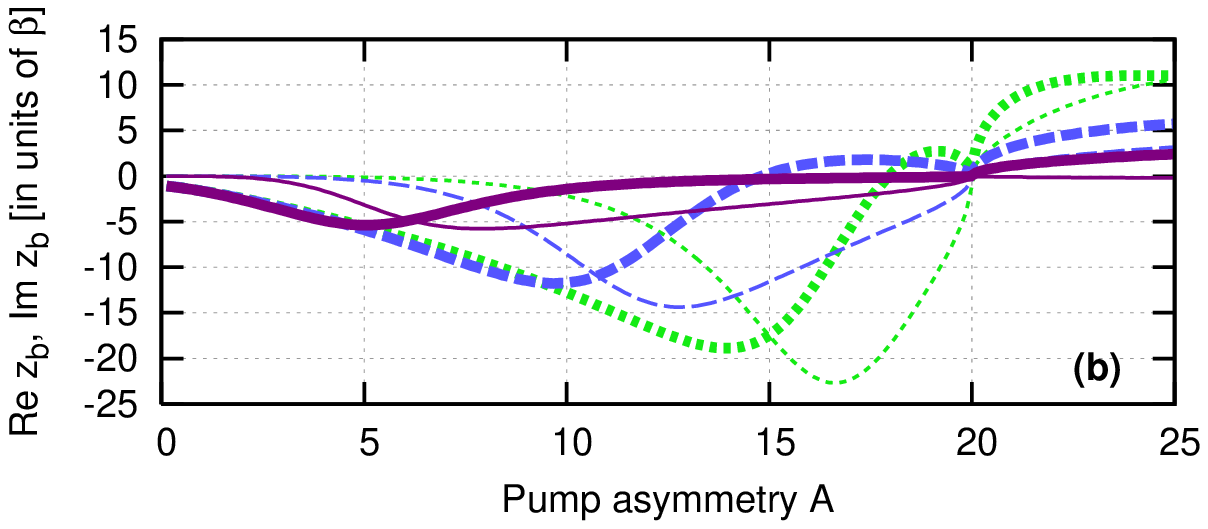}
\caption{ (color online) Real (thick) and imaginary (thin) part of the first few
  eigenvalues $z_1$ (continuous purple), $z_2$ (slashed blue), $z_3$
  (dotted green), for a lattice of $N=100$ (a) and $N=10$ (b) clouds
  of polarizability $\zeta=0.1$ each.
\label{fig:eigs} 
}
\end{center}
\end{figure}

The eigenvalue problem of a nonsymmetric real matrix is in
general not trivial. We have found, however, that a generalized
Fourier transformation with complex wavenumbers diagonalizes $\DD$
exactly. The analytical formulas for the eigenvectors
$\vv_b$ and eigenvalues $z_b$ of $\DD$, with $b=0,\ldots,N-1$, 
read
\begin{align}
\label{eq:eigenmodes}
[v_b]_j &= (\PP e^{2\pi i b})^{j/N},\\
\label{eq:eigenvalues}
z_b = {\beta}&\sqrt\PP \cos\Theta \left[
1 + \frac{4 \sqrt[N]{\PP} \sin^2\Theta}{( 
  \sqrt[N]{\PP} e^{i\pi b/N}  - e^{-i\pi b/N} )^2  }\right]^{-1}\!\!\!,
\end{align}
where $\beta= 8 k \zeta I_0 / c$ is related to the oscillation
frequency $\omega_0$ of a single cloud in a symmetric (incident laser
intensities $I_0 = I_1 = \epsilon_0\abs{E_0}^2 c/2$) trap by
$m\omega_0^2=\beta$.  Due to the pump asymmetry the eigenmodes
\eqref{eq:eigenmodes} of the lattice are complex, except for $b=0$,
which is a distorted center-of-mass mode and, if $N$ is even, $b=N/2$,
the density wave of highest wavenumber possible ($\pi/d$).  These two
modes are always stable, as $z_{N/2}<z_0<0$.  Since $\DD$ is real, all
other modes form conjugate pairs: $z_b=z_{N-b}^*$ and
$\vv_b=\vv_{N-b}^*$.  We briefly discuss the meaning of these
eigenmodes below.

Consider a pair of complex eigenvalues $z_b=z_{N-b}^*$
with $0<b<N/2$, and the corresponding eigenvectors $\vv_b =\vv_b^*$.
Both $\re(\vv_b)$ and $\im(\vv_b)$ describe density waves of
wavelength $N d /b$, modulated so their amplitude increases
towards the stronger pump.  Now time evolution by
(\ref{eq:atom_linear}) does not lead out of the subspace of $\RR^N$
spanned by these modes: for any superposition
$\boldsymbol{\xi}=p\re(\vv_b) + q \im(\vv_b)$ with $p,q\in\mathbb{R}$,
Eq.\eqref{eq:atom_linear} is equivalent to a single complex
homogeneous second-order linear differential equation, whose 
general solution is 
\begin{equation}
p+iq=c_+ e^{(\kappa_+ + i\omega_+)t}
+ c_- e^{(\kappa_- + i\omega_-)t}.
\end{equation}
Here $c_{\pm}=p_\pm + i q_\pm$ are arbitrary constants, and  
\begin{equation}
(\kappa_\pm + i\omega_\pm)= 
\frac{-\mu \pm \sqrt{\mu^2 + 4mz_b^*}}{2m},
\end{equation}
with $\kappa_-<\kappa_+$ to fix notation.  This corresponds to two
superimposed density waves of wavelength $Nd/b$, 
one copropagating with the stronger beam
($\omega_-<0$), and
one counterpropagating ($\omega_+>0$).  Their
phase velocities are given by $N d \abs{\omega_\pm} / (2\pi b)$.  The
copropagating wave is exponentially damped with constant $\kappa_-<0$,
but the counterpropagating wave can be either damped or amplified.
Thus, this pair of modes is stable if $\kappa_+<0$, which corresponds to
\begin{equation}
\label{eq:instability_beta}
m (\im\, z_b)^2 < -\mu^2 \re\, z_b.
\end{equation}

For symmetric pumping $\asym=0$, the matrix $\DD$ is symmetric, its 
eigenmodes \eqref{eq:eigenmodes}
are the Fourier components, and the eigenvalues \eqref{eq:eigenvalues}
are all real and negative, thus the lattice is stable.  Almost
all modes have the same frequency as a single trapped cloud,
$z_1=z_2=,\ldots,=z_N=-\beta$, except the center-of-mass mode, with
$z_0 = -\beta / (1+ N^2\zeta^2)$, which becomes soft if $N\to\infty$.

With the introduction of a pump asymmetry $\asym>0$, the eigenmodes
and the eigenvalues acquire imaginary parts, and as $\asym$ is
increased, the real parts of the eigenvalues turn positive one by one.
The first few eigenvalues are shown as functions of $\asym$ for two
examples in Fig.~\ref{fig:eigs}. In the ``strong collective
coupling'', $N\zeta\gg1$ limit (Fig.~\ref{fig:eigs} a), we observe
clearly separated resonances. In this limit, whenever $\pi-\Theta
\lesssim \pi/N$, we have $\sqrt[N]\PP\approx 1$, and the denominator
of \eqref{eq:eigenvalues} is approximately $1-\sin^2 \Theta/\sin^2
(\pi b/N)$, which, with Eq.\eqref{eq:theta}, places the resonance for
mode $b$ at $\asym \approx \asym_b = 2b\pi/(N\zeta)$.  We remark that
$\asym = 2 \pi/(N\zeta)$ fits the boundaries between the shaded green
areas of Fig.~\ref{fig:d_power} almost perfectly for $\asym<1$.
Outside of the strong collective coupling regime (Fig.~\ref{fig:eigs}
b), the resonances are not well resolved. It may even happen (as in
the plotted example) that mode $b=2$ becomes absolutely unstable
($\re\, z_2>0$) at lower $\asym$ than mode $b=1$. This causes the
``shoulder'' in the $N=10$ instability limit on
Fig.~\ref{fig:d_power}. At the critical asymmetry $\asym=2/\zeta=20$,
we have $\Theta=\pi/2$ and all eigenvalues are $0$; for $\asym>20$ all
modes are unstable.

\begin{figure}[!t]
\begin{center}
\includegraphics[width=8cm]{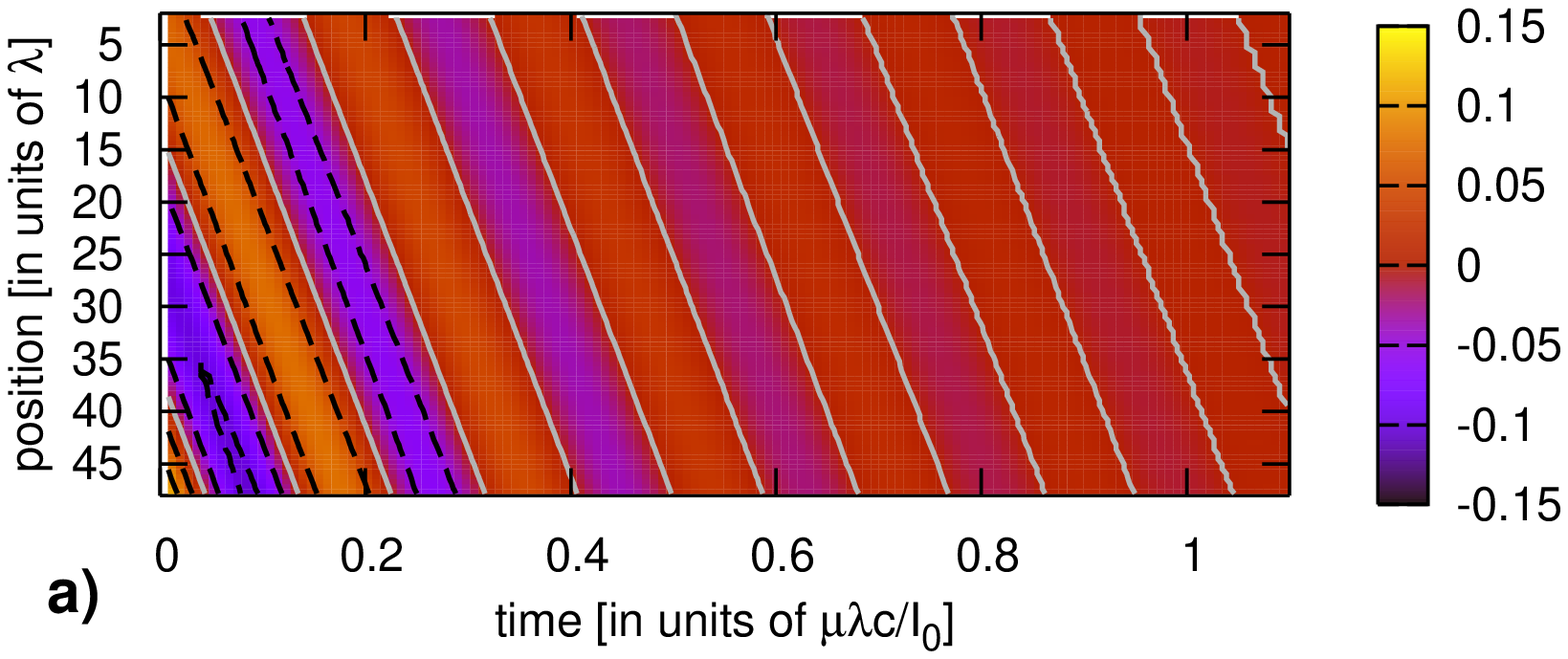}
\includegraphics[width=8cm]{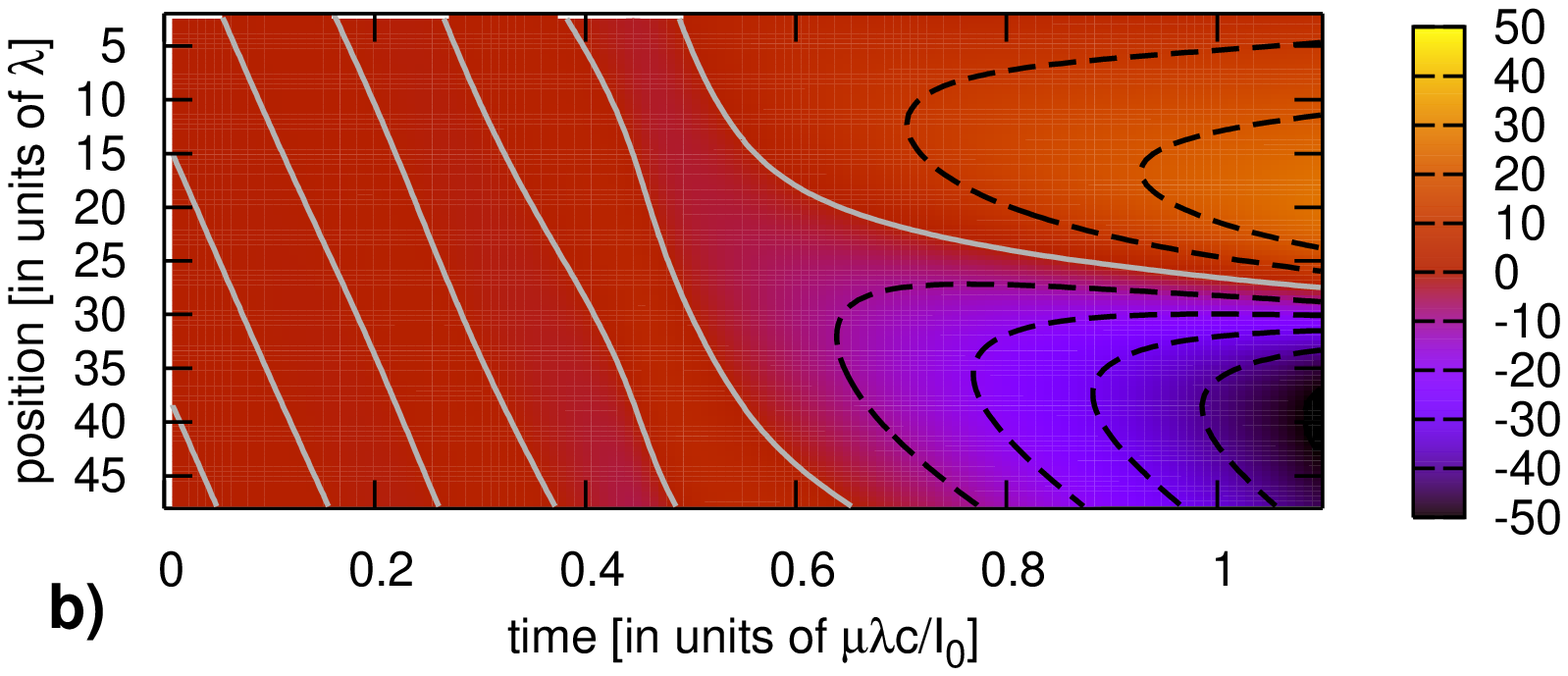}
\caption{(color online) Time dependence of position distortions $\xi$
  (in color coding, in units of $10^{-3}\lambda$) in an asymmetrically
  pumped overdamped optical lattice of $N=100$ clouds with
  polarizability $\zeta=0.1$, after excitation of mode $\re (\vv_1)$
  at $t=0$ with amplitude $10^{-3}\lambda$. The continuous grey
  contour line is $\xi=0$. In a), the system is subcritical: $\asym =
  0.632$, and $z_1/\beta = -0.55 -6.88 i$.  The excitation results in
  a density wave propagating towards the stronger beam, and dying out.
  In b), at supercritical asymmetry $\asym = 0.655$ the eigenvalue is
  $z_1/\beta = 1.48 -5.94 i$. The density wave is now amplified, and
  at $t\approx 0.5 \mu\lambda c /{I_0}\approx 2.5 \mu/\beta$ we leave the
  linear regime.  Then a local drop in the lattice constant develops
  at $x\approx 30\lambda$, which will result in two clouds coalescing,
  and eventually all particles will be pushed away by the stronger
  beam (not shown in figure).
\label{fig:waves} 
}
\end{center}
\end{figure}

A few remarks about the nature of these eigenmodes and the instability
are in order.  Two timescales govern the dynamics of the OL:
$\tosc=\sqrt{m/\abs{\re\, z_b}}$ of the oscillations and $\tdamp =
m/\mu$ of damping. For weak damping $\tosc\ll\tdamp$, modes with
nonzero $\im\, z_b$ are potentially unstable, but damping can restore
their stability, cf.~Eq.(\ref{eq:instability_beta}).  At the other
extreme, in the overdamped limit $\tdamp\ll\tosc$, the dynamics is
effectively first-order, and the copropagating mode disappears (is
``damped out''), for the counterpropagating mode we have $\omega_+ =
-\im\, z_b /\mu$, and $\kappa_+ = \re\, z_b /\mu$.  Even with
arbitrarily strong damping, the OL becomes unstable if $\re\, z_b>0$,
as the rhs of \eqref{eq:instability_beta} is negative. This ``absolute
instability'' is used to define the shaded areas of
Fig.~\ref{fig:d_power}.  We illustrate the dynamics close to the
absolute instability limit in Fig.~\ref{fig:waves}, showing the
results of numerical integration of Eq.~\eqref{eq:atom_dynamics} in
the overdamped regime near this limit.

Dynamical instabilities resulting from asymmetric pumping have been
observed in a far-detuned OL where atom--light interaction was
amplified by a ring cavity \cite{elsasser:033403}. In free space
near-resonant light has to be used (detunings of a few tens of atomic
linewidths seem realistic), and thus the influence of spontaneous
photons poses serious experimental limitations.  We checked via
simulation that the dissipative scattering force induces quantitative,
but no qualitative changes as long as $\abs{\im\, \zeta} <
\abs{\re\,\zeta}/100 $.  However, spontaneous emission also heats the
clouds, putting an upper limit on the timescale accessible by an
experiment, and complicating the very creation of the OL. One possible
way to circumvent the latter problem could be creating the OL at
larger detuning, where spontaneous heating is negligible, and then
continuously decreasing the detuning of the trap beams down to the
desired value.  As for the timescale of an experiment, we estimate
that, e.g., for a cold gas of Rb atoms in a dipole trap detuned by
$\Delta=-10\gamma\ldots -20\gamma$, forming $N=100$ disk-shaped
clouds, at pump power ratio $\PP=10$, the destabilization rate
$\kappa_+$ can exceed the heating rate by orders of magnitude if the
surface density of the clouds is $\eta>1/(2\lambda^2)$.


We acknowledge funding from the Austrian Science Foundation 
(Contract Nos.~P17709 and S1512), and
the National Scientific Fund of Hungary (NF68736, T043079 and T049234).

\end{document}